# A Convolutional Neural Networks Denoising Approach for Salt and Pepper Noise

Bo Fu, Xiao-Yang Zhao, Yi Li, Xiang-Hai Wang and Yong-Gong Ren

School of Computer and Information Technology, Liaoning Normal University, China
fubo@lnnu.edu.cn

**Abstract.** The salt and pepper noise, especially the one with extremely high percentage of impulses, brings a significant challenge to image denoising. In this paper, we propose a non-local switching filter convolutional neural network denoising algorithm, named NLSF-CNN, for salt and pepper noise. As its name suggested, our NLSF-CNN consists of two steps, i.e., a NLSF processing step and a CNN training step. First, we develop a NLSF pre-processing step for noisy images using non-local information. Then, the pre-processed images are divided into patches and used for CNN training, leading to a CNN denoising model for future noisy images. We conduct a number of experiments to evaluate the effectiveness of NLSF-CNN. Experimental results show that NLSF-CNN outperforms the state-of-the-art denoising algorithms with a few training images.

## 1      Introduction

Images are always polluted by noise during image acquisition and transmission, resulting in low-quality images.  The removal of such noise is a must before subsequent image analysis tasks. To achieve this, researchers have proposed many denoising algorithms [1-7], where the representative ones include non-local means (NLM) [1] and block-matching 3D filtering (BM3D) [2].In this paper, we focus on the salt and pepper noise, where pixels exhibit maximal or minimal values with some pre-defined probability. To our knowledge, the most commonly used denoising algorithm for salt and pepper noise is the nonlinear filter family, such as the median filter, with noise detecting step [10-15]. Empirical results reported in previous studies show that these traditional algorithms are effective for salt and pepper noise to some extent [16, 22, 30-33]. Wang et al tried to apply hash retrieval and low rank methods to compute similarity and code[34-42]. However, such traditional algorithms are heavily dependent on local information of images. Thus, they often perform worse when facing high density noise without sufficient exact information. This is problematic in real-world image applications.

In view of the above problems, the motivations of our research have three point: First point, we design a switching templates to avoid noise disturb in the process of measuring similarity. Second, we extract repairable information in a non-local region in-stead of a local patch. Based on two points above, we proposed a navel non-local switching filter (NLSF) as pre-processing part. Because NLSF can remove most of noise, so CNN can more effective discover the non-linear relationship between lost details in noisy image and original image. The third point, we proposed a convolutional neural networks denoising algorithm embracing our NLSF and CNN named NLSF-CNN.

Under the framework of NLSF-CNN, we optimize the repair information analysis by expanding the search scope to a non-local scope, and build a switching templates to avoid noise disturbing in the process of calculating weights. A set of patches filtered by NLSF are input CNN, this mapping of neural networks will be learnt from a filtered image to the ground-truth image. In the front part of NLSF-CNN, we use our NLSF to pre-process noisy image for making sure the learning process is not in a high noise environment. This step will bring great benefits to subsequent networks learning. At the same time, CNN with multi-layer structure is effective on increasing the capacity and flexibility for exploiting details which lost in previous filtering process.

The contributions of our work are summarized as following: First, we propose a new Salt and Pepper noise image denoising algorithm based on convolutional neural net-works. In contrast to traditional CNNs for denoising which directly input noisy images and estimate clean image, our network is added a pre-filtering step(NLSF) and learns clean image from a mapping between filtered image(not noisy image) and the corresponding ground-truth image. Second, we propose a new non-local switching filter. It can be well executed in the pre-processing part of the network. This pre-processing step can reduce the interference of high-intensity noise in deep learning process. A series of experiments are performed and our algorithm obtain a higher PSNR score than other methods. The visual results indicate that our algorithm can better supplement details.

The remainder of this paper is organized as follows. In section 2, we introduce patch model and networks structure firstly, and introduce the denoising process of our neural network. In Section 3, we test several simulation experiments and compare with state-of-the-art methods. In Section 4, we conclude and prospect.

## 2 Related Work

With the continuous development of deep learning theory, more and more image processing problems were tried by variety types of networks and achieved good results. In considering existing problems mentioned above, some scholars apply more powerful deep learning methods to the image denoising field. V. Jain firstly added a CNN model to denoising method in 2008[17].Their algorithm not only achieves better results than the traditional wavelet and hidden Markov models, but also shows that a particular form of CNN can be regarded as an approximation of the result of Markov model inference for image denoising. At the same time, CNN model can avoid the computational difficulty of Markov model in probability learning and inference. Rectifier Linear Unit [18], batch normalization [19] are also proposed comparing to the traditional methods [21]. In 2012, Xie used stacked denoising auto-encoder for image de-noising and image restoration [23]. In 2017, Zhang proposed a deeper CNN network called DnCNN [20].

These methods have made great progress in image denoising field, but they are all aimed at removal of Gauss white noise. At present, there is still lacking of an effective network structure between an image of salt and pepper noise pollution and the corresponding ground-truth image.

## 3 Convolutional neural networks for Salt and Pepper noise image denoising

In this section, we briefly review image denoising with salt and pepper noise, and then introduce the proposed non-local switching filter convolutional neural network denoising algorithm, named NLSF-CNN.

### 3.1 Denoising with Salt and Pepper Noise

In order to facilitate later use, we first named some commonly used variables. The input noise image is cut into a series of overlapping patches, the generation process is shown in Figure 2. For a patch P, let (i, j) be the coordinates of a pixel, $I_{(i,j)}$ is the value of pixel (i,j). We first detect salt and pepper noise pixels by using a threshold $\delta$. For each pixel $I_{(i,j)}$, the noise detection follows the following rule:

$$I_{(i,j)} = \begin{cases} \infty & if\ I_{(i,j)} \in [0,\delta)\ or\ (255-\delta, 255] \\ I_{(i,j)} & otherwise \end{cases} \quad (1)$$

where $\infty$ denotes the salt and pepper noise[1].

Usually, the values of salt and pepper noise distribute both ends of the gray scale range, i.e. $[0,\delta)$ or $(255-\delta, 255]$. Here, $\delta$ represents a threshold. Because most noise values are concentrated at both ends of the gray scale, the pixels which gray-value located in the middle of the gray-scope are most likely to be normal pixels Probability. So, classical non-linear filter usually use median value to repair noise. Of course, many advanced algorithms have been followed up concerning more valid pixel. However, these methods are faced with a key problem, i.e. repair steps are often dependent on analysis and measurement, but these analysis and measurement methods will be disturbed by noise. So, How to extract details more effectively without noise interference is still a problem for scholars to solve.

---

[1] Here, we empirically set $\delta$ to 1.

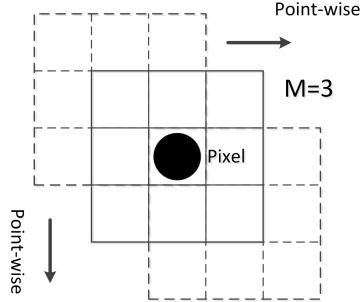

**Fig. 1.** Patches generation of size L=3

### 3.2 NLSF-CNN

We introduce the proposed NLSF-CNN denoising algorithm for salt and pepper noise. In NLSF-CNN, we train a patch-based CNN model over training images with artificial noise, and then use the model as a denoiser for future noisy images. NLSF-CNN consists of two steps, including a pre-processing step and a CNN training step. In the pre-processing step, we first detect the noisy pixels of salt and pepper and then smooth them using a non-local switching filter method. In the CNN training step, the pre-processed images are divided into overlapping patches. We use these patches as the input for CNN, and train the optimal parameters of CNN.

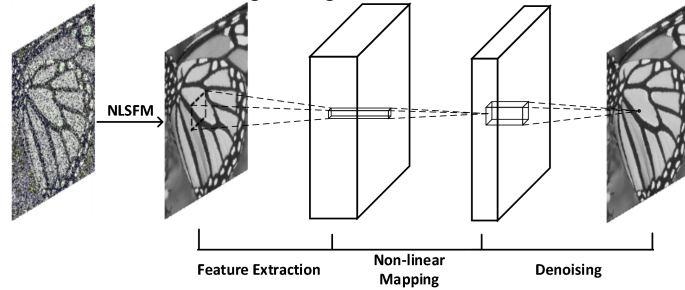

**Fig. 2.** Network structure

We now present details of each step in NLSF-CNN, and the selection of parameters is discussed in the 4.1 section.

**Pre-processing Step.** Given training images with artificial salt and pepper noise, we pre-process noisy pixels before feed them into CNN training. This is because the salt and pepper noise is a pure noise, which is harmful to CNN training. In this work, we propose a non-local switching filter method in the pre-processing step.

Then, we use non-local information to remove noisy pixels one by one. For each noisy pixel, generate $R$ point-wise patches of size $L$ around it. We replace the noisy pixel with a weighted sum of medians of these $R$ patches:

$$\tilde{I}_{(i,j)} = \sum_{\forall (k,l) \in \Omega_n} w_{(k,l)} \cdot I_{(i,j)} \quad (2)$$

where $w_{(k,l)}$ is the weight of patch $p_k$ for $p_l$, and $0 < w_{(k,l)} < 1$. The weight value is depended on the Euclidian distance between any patch and the patch centered by the current noisy pixel:

$$w_{(k,l)} = \frac{s_{(k,l)}}{\sum_1^n s_{(k,l)}} \quad (3)$$

$$s_{(k,l)} = \frac{1}{e^{\frac{\|P_k - P_l\|^2}{\sigma_n}} \ln \sigma_n} \quad (4)$$

To obtain more accurate distances between patches, we replace all noisy pixels with the mean of patches during distance measuring.

**Generating Non-linear mapping and Reconstruction.** So far, all patches in set P are pre-filtered. We input these patches into our networks. Our networks own three layers, in each layers, we define a set of filters and operators to generate mappings.

In the first layer, we represent these patches by a set of bases. The operation can by describe as follows:

$$F_1(P) = \max(0, W_1 * P + B_1) \quad (5)$$

where W and B represent the bases filter set and biases respectively, size of filter is $f_1$. We denote the number of bases filters is n, and filter's size is same to size of patch. So, each patch is applied n convolutions. The convolutional result of each patch is corresponding an n-dimensional feature map. Then we apply the Rectified Linear Unit (ReLU, max(0,x)) [24] on the filter responses to obtain more non-linearity.

Based on an n-dimensional feature extracted in first layer, we map each of these n-dimensional features into a new n-dimensional feature. The operation of the second layer is:

$$F_2(P) = \max(0, W_2 * F_1(P) + B_2) \quad (6)$$

Here $W_2$ is of a size n×1×1×n, this is equivalent to applying filters which have a spatial support 1×1, and $B_2$ is biases. Then we still apply the Rectified Linear Unit. The second layer can increase the non-linearity between pre-filtered patch and ground-truth.

In the third layer, we define a convolutional layer to predict enhancing patches, and reconstruct them as a result image. The operation of the third layer is:

$$F_3(P) = \max(0, W_3 * F_2(P) + B_3) \quad (7)$$

here $W_3$ is of a set of filter, and $B_3$ is biases.

In the process of learning mapping, we estimate mapping function F of parameters $\Theta = \{W_1, W_2, W_3, B_1, B_2, B_3\}$. This is a processing of achieving minimizing the loss between the result images F(Y;Θ) and the corresponding ground truth. We use Mean Squared Error (MSE) as the loss function:

$$\text{MSE} = \frac{1}{MN} \sum_{i=1}^{M} \sum_{j=1}^{N} \left[ I_{(i,j)} - \tilde{I}_{(i,j)} \right] \tag{8}$$

We give NLSF-CNN a set of noisy images and their ground-truth images and obtain a mapping which describing their nonlinear relationship.

### 3.3   Testing NLSF-CNN

For each noisy image, we first use the NLSF method to compute a pre-processed version of this image. Then, we divide it into patches with legal size and feed them into the pre-trained CNN model for clean outputting patches. Finally, we integrate these clean patches, leading to the clean image.

## 4   Experiment

This section shows empirical results. We compare NLSF-CNN against the sate-of-the-art baseline algorithms, and also empirically evaluate its crucial parameters.

### 4.1   Experimental Setting

***Dateset*** We employ a public training set from [25-28], which contains 91 training images.

During testing, we use two different test sets to evaluate our algorithm. (1) A small test set, i.e., the standard test images, contains 11 commonly used images for denoising, e.g., "Lena", "Baboon" and "Pepper". (2) A big test set, i.e., Berkeley segmentation dataset (BSD300)[2], contains 300 images.

For our NLSF-CNN, the parameters are set as follows: In the process of training the networks, we set f1=9, f3=5, n1=64, n2=32 in main evaluations.

***Baseline algorithm*** We use four existing denoising algorithms as baselines, including three traditional algorithms and a neural network based one.

The traditional algorithms include decision based algorithm (DBA) [10][3], patch-based approach to remove impulse-Gaussian noise from images (PARIGI) [22] [4] and adaptive switching non-local filter[2] (NASNLM) [24].

The neural network based algorithm is the multi-layer perception (MLP) proposed in [21]. Here, we use the pre-trained MLP model provided by its authors. For fair comparisons, we perform our NLSF pre-processing step to test noisy images before applying the pre-trained MLP model[5], referring to as NLSF-MLP.

---

[2] https://www2.eecs.berkeley.edu/Research/Projects/CS/vision/grouping/segbench/BSDS300/html/dataset/images.html
[3] https://github.com/gkh178/noise-adaptive-switching-non-local-means/tree/master/NASNLM
[4] http://www.ipol.im/pub/art/2016/161/
[5] https://github.com/urbste/MLPnP_matlab_toolbox

Besides, we use the proposed NLSF pre-processing method as a supplement baseline.

For our NLSF-CNN, the parameters are empirically set as follows: (1) In the NLSF pre-processing step, the patch size is set to 3 when the noise density is lower than 30%, otherwise 5. (2) In the CNN training step, the size of input patch is set to 64.

***Evaluation Metric*** The peak signal-to-noise ratio (PSNR) is adopted to measure the objective performance of our algorithm. PSNR is defined as follows.

$$\text{PSNR} = 10\log_{10}(\frac{255^2}{MSE}) \quad (9)$$

Here, MSE is Mean Square Error, it is defined in Formula 8.

### 4.2  Performance Comparison

This subsection shows the evaluation results. We examine our NLSF–CNN with different noise densities, i.e., 30%, 50% and 70%.

Table 1. PSNR scores of standard test images.

| Test Image | level | DBA | NASNLM | PARIGI | NLSF | NLSF-MLP | NLSF-CNN |
|---|---|---|---|---|---|---|---|
| Lena | 30% | 34.42 | 28.09 | 33.90 | 34.20 | 30.80 | **35.38** |
|  | 50% | 30.11 | 26.15 | 29.91 | 30.12 | 29.28 | **32.55** |
|  | 70% | 25.84 | 25.97 | 25.22 | 25.79 | 27.63 | **30.18** |
| Bridge | 30% | 28.07 | 23.68 | 25.19 | 28.21 | 25.19 | **28.71** |
|  | 50% | 24.24 | 22.91 | 22.61 | 24.45 | 23.86 | **26.01** |
|  | 70% | 21.12 | 22.63 | 20.06 | 21.02 | 22.61 | **24.11** |
| Girl | 30% | 29.41 | 20.61 | 29.74 | 32.88 | 29.64 | **33.47** |
|  | 50% | 27.47 | 16.69 | 27.25 | 29.66 | 28.28 | **30.92** |
|  | 70% | 24.99 | 16.32 | 24.29 | 26.33 | 26.90 | **29.06** |
| pepper | 30% | 26.85 | 22.38 | 28.88 | 32.27 | 30.01 | **32.99** |
|  | 50% | 25.27 | 21.82 | 25.44 | 27.99 | 28.57 | **30.23** |
|  | 70% | 22.11 | 21.58 | 21.46 | 23.04 | 27.04 | **27.70** |
| Average of 11 images | 30% | 31.79 | 27.07 | 30.86 | 32.28 | 29.77 | **33.35** |
|  | 50% | 28.27 | 26.38 | 27.47 | 29.28 | 28.09 | **31.34** |
|  | 70% | 24.38 | 26.98 | 23.87 | 25.09 | 26.36 | **29.15** |

Table 1 shows the PSNR results on the standard test set, including four example standard test images, e.g., "Lena" and "Bridge", and average results of all 11 images of this test set. Overall, we can observe that our NLSF-CNN outperforms baseline algorithms in all settings. Some detailed observations are made. (1) Comparing with the three traditional denoising algorithms, the gain of NLSF-CNN is significant. For example, the PSNR scores of NLSF-CNN are about 5~7 higher than those of NASNLM on "Lena" and about 4~6 higher than those of PARIGI on "Pepper". (2) Comparing with NLSF-MLP, our NLSF-CNN also performs better. This indicates that NLSF-CNN is much more practical, since we use a very small training set, i.e., 91 training images only, but MLP is trained on hundreds of thousands of images. (3) We can observe that the gain of NLSF-CNN becomes more significant with high noise densities. For example, the average PSNR of NLSF-CNN is about 3~6 higher than those of baseline algorithms when the noise density reaches 70%. This implies that NLSF-CNN is more robust and practical in real world image applications.

We also perform comparative methods on BSD300, and put average PSNR scores of 300 images in Table2.

**Table 2.** The average results of PSNR (dB) on the BSD300 dataset.

| Test Image | Noise level | DBA | NASNLM | NLSF | NLSF-MLP | NLSF-CNN |
|---|---|---|---|---|---|---|
| BSD300 | 30% | 29.92 | 25.74 | 30.01 | 29.77 | **30.87** |
| BSD300 | 50% | 26.32 | 24.5 | 26.25 | 26.19 | **27.84** |
| BSD300 | 70% | 22.81 | 24.65 | 22.85 | 24.72 | **25.35** |

Table 2 shows the average PSNR results on BSD300 test set with different noise densities. We also observe that NLSF-CNN performs better than baseline algorithms, especially for higher noise density. This further indicates that our NLSF-CNN is more effective for the salt and pepper denoising. Besides, NLSF-CNN outperforms NLSF-MLP, but uses much less training images, saving many training time.

The visual results are also compared in Figure 3, and Figure 4. In Figure 3, we compared original clean image, 50% noisy image, and results images of DBA, PARIGI, NASNLM, NLSF-MLP, NLSF and NLSF-CNN.

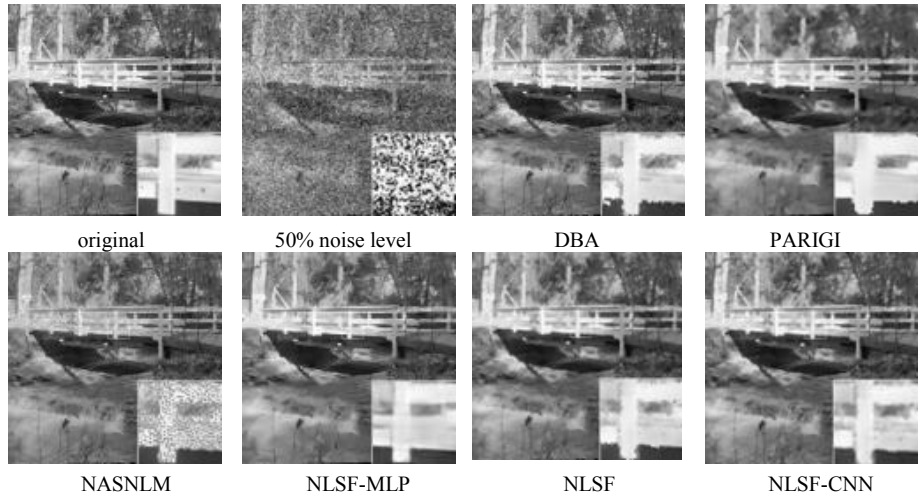

original     50% noise level     DBA     PARIGI

NASNLM     NLSF-MLP     NLSF     NLSF-CNN

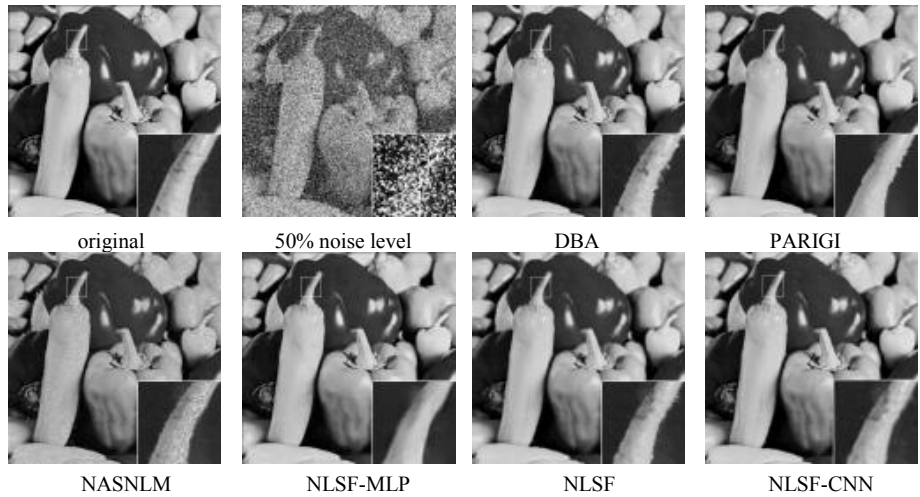

**Fig. 3.** The test results of all comparative algorithms

The zoomed details are added to the lower right corner of these images. In 50% noise density, DBA and NASNLM are not completely removing noise, and some tiny artifacts consist in result of NLSF. NLSF-MLP over smooth details, for example it loses two black nails in image of boat, and loses some tiny black shadow on the stem of a pepper. It can be seen that NLSF-CNN not only removes noise but also obtains satisfactory visual effects and preserves the details.

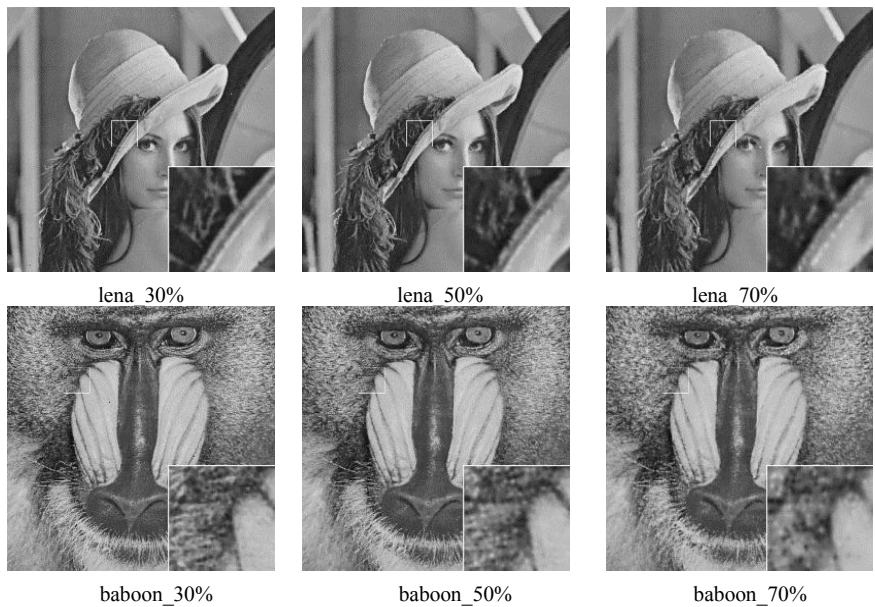

**Fig. 4.** Some results and details by NLSF-CNN

In Figure 4, we performed our algorithm in different texture types of images in 30%, 50%. 70% noise density levels. It can be seen that satisfactory visual effects were obtained, especially the edges were well protected.

### 4.3    Parameter evaluation

In this subsection, we empirically evaluate the patch size in the NLSF preprocessing step. To achieve this, we show the PSNR results of different patch sizes, i.e., 3, 5 and 7, with different noise densities. Figure 5 shows the results. We can see that the PSNR scores of size 3*3 are higher than others when the noise density is lower than 30%, and those of size 5*5 are better for higher noise densities. The possible reason is that small patches may contain insufficient local information for high noise densities.

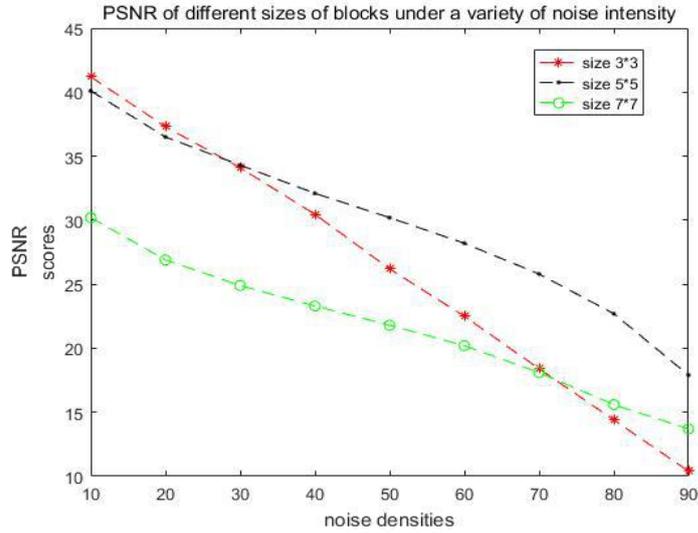

**Fig. 5.** Performance choose different size of patch in various noise intensities levels

In this subsection, we empirically evaluate the patch size in the NLSF pre-processing step. To achieve this, we show the PSNR results of different patch sizes, i.e., 3, 5 and 7, with different noise densities. Figure 5 shows the results. We can see that the PSNR scores of size 3*3 are higher than others when the noise density is lower than 30%, and those of size 5*5 are better for higher noise densities. The possible reason is that small patches may contain insufficient local information for high noise densities.

## 5    Conclusions

In this paper, we propose a convolutional neural networks denoising algorithm, namely NLSF-CNN, for Salt and Pepper noise. New non-local switching median filter, learning algorithm and neural networks architecture are embraced in our image de-noising algorithm. There are three aspects of the innovation of our algorithm. First, we define a patch model to describe the local image contaminated by noise. Second, we design a switching median filter based self-similarity model above. Third, we designed a CNN structure by introducing our switching median filter, and then the mapping is learning from filtered image to target image.

Simulation results demonstrate that the proposed algorithm implements effectively in various salt and pepper noise densities. It can be seen from Table1, Table 2, Figure 3 and figure 4 that our algorithm obtain satisfactory visual effects and higher PSNR scores only use 91 images to train networks. The reason that proposed algorithm has a better effect is our networks can more efficient mining details and add them to the NLSF result. These results prove that our algorithm is effective and meaningful.

## 6    Acknowledgement

This work is supported by the National Natural Science Foundation of China (NSFC) No. 61702246, Liaoning Province of China General Project of Scientific Research No. L2015285, Liaoning Province of China Doctoral Research Start-Up Fund No. 201601243.

...